\pdfoutput=1
\documentclass[a4paper,12pt]{article}
\usepackage{amsmath}
\usepackage{amssymb}
\usepackage{amsfonts}
\usepackage{graphicx}
\usepackage{hhline}
\usepackage{xfrac}
\usepackage{bbold}

\usepackage[utf8]{inputenc}

\usepackage[square,numbers,comma,sort&compress]{natbib}
\usepackage[colorlinks=true,citecolor=blue,linkcolor=purple,urlcolor=purple]{hyperref}
\usepackage{geometry}
\usepackage[usenames,dvipsnames,svgnames,table]{xcolor}
\usepackage{booktabs}
\usepackage{slashed}

\def\varabstract{ }
\def\varkeywords{ }
\def\vararxivnumber{ }
\def\vartitle{ }
\def\varsubtitle{ }
\renewcommand{\title}[1]{\gdef\vartitle{#1}}

\renewcommand{\abstract}[1]{\gdef\varabstract{#1}}
\newcommand{\keywords}[1]{\gdef\varkeywords{#1}}

\newtoks\authtoks
\renewcommand{\author}[2][]{%
	\authtoks=\expandafter{\the\authtoks#2$^{#1}$\ }%
}
\newtoks\affiltoks
\newcommand{\affiliation}[2][]{%
    \affiltoks=\expandafter{\the\affiltoks{\item[$^{#1}$]#2}}%
}
\newtoks\emailtoks\newcounter{emailcounter}%
\newcommand{\emailAdd}[1]{%
\ifnum\theemailcounter>0\emailtoks=\expandafter{\the\emailtoks, \typeemail{#1}}%
\else\emailtoks=\expandafter{\typeemail{#1}}%
\fi
\stepcounter{emailcounter}}
\newcommand{\typeemail}[1]{\href{mailto:#1}{\tt #1}}

\renewcommand\maketitle{
	\newgeometry{margin=2cm}
	\pagestyle{empty}\setcounter{page}{0}
	{\huge\flushleft\sffamily\bfseries\vartitle\\\Large\varsubtitle\par}
\vskip6ex
{\large\bfseries\raggedright\sffamily\the\authtoks\par}
\vskip2ex
\begin{list}{}{%
\setlength{\leftmargin}{0.28cm}%
\setlength{\labelsep}{0pt}%
\setlength{\itemsep}{-3pt}%
\setlength{\topsep}{-\parskip}}
\itshape\small%
\the\affiltoks
\end{list}
\vskip2ex
\noindent\hspace{0.28cm}\begin{minipage}[l]{.9\textwidth}
\begin{flushleft}
\textit{E-mail:} \the\emailtoks
\end{flushleft}
\end{minipage}
\vskip5ex
\noindent{\renewcommand\baselinestretch{.9}\textsc{Abstract:}}\ \varabstract
\vskip5ex
\if!\varkeywords!\else\noindent{\textsc{Keywords:}}\ \varkeywords \vskip2ex\fi
\newpage
\restoregeometry
\pagestyle{plain}

\setcounter{footnote}{0}
}

\usepackage[square,numbers,comma,sort&compress]{natbib}

\definecolor{MS}{rgb}{0,0,1}

\usepackage{ifthen}

	\newcommand{\barlimc}[7]{
  \pgfmathparse{\mypos+0.3}
  \edef\mypos{\pgfmathresult}
		\node[left,scale=0.6] at (0,\mypos) {#1};
		\pgfmathparse{#3 > 5 ? 1 : 0}
		\ifthenelse{\pgfmathresult=1}{
			\fill[#2] ($(0,\mypos)+(0,-0.1)$) rectangle +(5,0.2);
			\fill[white] ($(0,\mypos)+(3.5,-0.1)$) rectangle +(0.3,0.2);
			\draw[decoration={zigzag},decorate,#2,very thick] (3.4,\mypos) to +(0.5,0);
			\node[left,scale=0.6] at (5,\mypos) {#3};
			}{
			\fill[#2] ($(0,\mypos)+(0,-0.1)$) rectangle +(#3,0.2);
			\node[left,scale=0.6] at (#3,\mypos) {#3};
		}
		\fill[#4] ($(0,\mypos)+(0,-0.1)$) rectangle +(#5,0.2);
		\node[left,scale=0.6] at (#5,\mypos) {#5};
		\fill[#6] ($(0,\mypos)+(0,-0.1)$) rectangle +(#7,0.2);
		\pgfmathparse{#7 <0.3 ? 1 : 0}
		\ifthenelse{\pgfmathresult=1}{
			\node[right,scale=0.6] at (0,\mypos) {#7};
		}{
		\node[left,scale=0.6] at (#7,\mypos) {#7};
	}
}

\title{LHC constraints on $W^\prime,~Z^\prime$ that couple mainly to third generation fermions}

\author[1]{Alper Hayreter}\emailAdd{alper.hayreter@ozyegin.edu.tr}
\author[2,3,4]{Xiao-Gang He,}\emailAdd{hexg@phys.ntu.edu.tw}
\author[5]{ and German Valencia}\emailAdd{german.valencia@monash.edu}

\affiliation[1]{Department of Natural and Mathematical Sciences, Ozyegin University, 34794 Istanbul Turkey.}
\affiliation[2]{Tsung-Dao Lee Institute $\&$ Department of Physics and Astronomy, SKLPPC,
Shanghai Jiao Tong University, 800 Dongchuan Rd., Minhang, Shanghai 200240, China}
\affiliation[3]{Department of Physics, National Taiwan University,\\
No.\,\,1, Sec.\,\,4, Roosevelt Rd., Taipei 106, Taiwan}
\affiliation[4]{Physics Division, National Center for Theoretical Sciences,\\
No.\,\,101, Sec.\,\,2, Kuang Fu Rd., Hsinchu 300, Taiwan}
\affiliation[5]{School of Physics and Astronomy, Monash University, Melbourne VIC-3800, Australia}

\abstract{We use the results of CMS and ATLAS searches for resonances that decay to $\tau\nu$ or $tb$ and $\tau^+\tau^-$ or $t\bar t$ final states to constrain the parameters of non-universal $W^\prime$ and $Z^\prime$ gauge bosons that couple preferentially to the third generation. For the former we consider production from $c\bar{b}$ annihilation and find very weak constraints on the strength of the interaction and only for the mass range between 800 and 1100 GeV  from the $pp \to \tau_h p_T^{\rm miss}$ channel. The constraints on the latter are much stronger and arise  from both $t\bar t$ and $\tau^+\tau^-$ production. Treated separately, we find that the weak constraints on the $W^\prime$ still permit an explanation of the $R(D^{(\star)})$ anomalies with a light sterile neutrino whereas the stronger constraints on the $Z^\prime$ exclude significant light sterile neutrino contributions to the $K \to \pi \nu \bar\nu$ rates. Within specific models the masses of  $W^\prime$ and $Z^\prime$ are of course related and we briefly discuss the consequences. }


\begin{document}
\baselineskip=17pt \parskip=5pt

\maketitle

{\hypersetup{linkcolor=black}
  \tableofcontents}

\newpage

\section{Introduction}

The CMS and ATLAS experiments have both searched for exotic resonances that single out the third generation. In particular CMS has reported a result from an integrated luminosity of 35.9~fb$^{-1}$ at $\sqrt{S}=13$~TeV on searches for a $W^\prime$ that decays into a tau-lepton and a neutrino \cite{Sirunyan:2018lbg} while ATLAS has reported results for the same channel with 36.1~fb$^{-1}$ at $\sqrt{S}=13$~TeV \cite{Aaboud:2018vgh}. In both cases the analysis was carried out with benchmark models for the $W^\prime$ in which production through $u\bar{d}$ annihilation is assumed. This is inadequate for truly non-universal models where the couplings to the first generation quarks can be suppressed by many orders of magnitude. We reinterpret these searches by comparing the experimental results to cross-sections  obtained from $c\bar{b}$ annihilation production of the $W^\prime$. ATLAS has also reported  a search in the top-bottom final state with 36.1~fb$^{-1}$ at $\sqrt{S}=13$~TeV \cite{Aaboud:2018jux} which is not yet competitive with the $\tau\nu$ channel.

In the same manner, searches for non-universal $Z^\prime$ bosons have also been reported in the $\tau^+\tau^-$ channel by CMS with 2.2~fb$^{-1}$ at $\sqrt{S}=13$~TeV \cite{Khachatryan:2016qkc} and in the $t\bar t$ channel by ATLAS with 36.1~fb$^{-1}$ at $\sqrt{S}=13$~TeV \cite{Aaboud:2019roo}. The models studied in these cases also assumed couplings of the  $Z^\prime$ to first generation fermions and we reinterpret those results for models in which the  $Z^\prime$ is dominantly produced from $b\bar b$ annihilation.

Our study is motivated by a possible explanation for the so-called charged B anomalies. Measurements of $R(D)$  \cite{Lees:2012xj,Lees:2013uzd,Huschle:2015rga} and $R(D^\star)$  \cite{Lees:2012xj,Lees:2013uzd,Huschle:2015rga,Sato:2016svk,Hirose:2016wfn,Aaij:2015yra} show hints for a deviation from the SM. These observables are defined as
\begin{eqnarray}
R(D)&=&\frac{B(\bar{B}\to D\tau^-\bar\nu_\tau)}{B(\bar{B}\to D\ell^-\bar\nu_\ell)},~{\ell=\mu,e}\nonumber\\
R(D^\star)&=&\frac{B(\bar{B}\to D^\star\tau^-\bar\nu_\tau)}{B(\bar{B}\to D^\star\ell^-\bar\nu_\ell)}.
\end{eqnarray}
In both cases the measurements are higher than the SM with a combined significance of about $3\sigma$. Many new physics models have been proposed as possible explanations. One class of such models accomplishes these enhancements with a new right-handed neutrino and a (right-handed) $W^\prime$ \cite{He:2012zp,He:2018uey,Greljo:2018ogz,Robinson:2018gza,Babu:2018vrl,Gomez:2019xfw}. The couplings of the $W^\prime$ are non-universal, preferring the third generation as suggested by the results for $R(D^{(\star)})$. These studies suggest that a viable explanation within these models is possible with a $W^\prime$ mass near 1~TeV, that therefore could be studied by the LHC. The authors of Ref.~ \cite{Greljo:2018tzh} have argued using an effective Lagrangian that the operator relevant for this explanation of  $R(D^{(\star)})$, would result in an enhanced production of mono-tau events that is ruled out by CMS measurements \cite{Sirunyan:2018lbg}. In this paper we reexamine that result for a specific model, finding that this explanation of the charged B anomalies is not yet ruled out by LHC data. 

 For a benchmark non-universal right-handed $W^\prime$, we rely on an $SU(2)\times SU(2)$ model that we have studied at length before \cite{He:2002ha,He:2003qv}.  The model also contains a $Z^\prime$ which can have interesting consequences for rare flavour processes such as $K\to \pi \nu\bar\nu$. Our paper is organised as follows: in Section 2 we review the details of the model that are necessary for this study and summarise known constraints. In Section 3 we confront the $W^\prime$ and $Z^\prime$ in the parameter region of interest for the flavour results with available LHC searches. In Section 4 we revisit the viability of these scenarios as sources for large deviations in flavour observables, and in Section 5 we conclude. Taking the $W^\prime$ on its own, we find that in combination with a light sterile neutrino, this is still a viable explanation for the $R(D^{(\star)})$ anomalies. On the other hand, the constraints on the $Z^\prime$ rule out significant enhancements to $K\to \pi \nu \bar\nu$ modes through the addition of light sterile neutrinos that couple to the $Z^\prime$. We comment on connections between the two sets of bounds within the context of a specific model.

\section{The model}

Our starting point will be the non-universal LR model of \cite{He:2002ha,He:2003qv}. To single out the third generation we augment the SM gauge group with a second $SU(2)$ under which only the third generation right handed fermions are charged. The gauge group is then $SU(3)_C\times SU(2)_L\times SU(2)_R \times U(1)_{B-L}$ with gauge coupling constants $g_3$, $g_L$, $g_R$ and $g$, respectively. In the weak interaction basis, the first two generations of quarks $Q_L^{1,2}$, $U_R^{1,2}$, $D_R^{1,2}$ transform as $(3,2,1)(1/3)$, $(3,1,1)(4/3)$ and $(3,1,1)(-2/3)$, and the leptons $L_L^{1,2}$, $E_R^{1,2}$ transform as $(1,2,1)(-1)$ and $(1,1,1)(-2)$. The third generation, on the other hand, transforms as $Q_L^3\;(3,2,1)(1/3)$, $Q^3_R\;(3,1,2)(1/3)$, $L^3_L\;(1,2,1)(-1)$ and $L^3_R\;(1,1,2)(-1)$. These assignments provide universality violation. The third family has, in addition to the SM fermions, a sterile neutrino which we denote by $N_R$. It appears as the partner of the $\tau_R$ in the right-handed doublet $L^3_R$.

Since the new right-handed gauge bosons must be heavier than the $W$ and $Z$, we separate the symmetry breaking scales of $SU(2)_L$ and $SU(2)_R$,  introducing two 
Higgs multiplets $H_L\; (1,2,1)(-1)$ and $H_R\;(1,1,2)(-1)$ with respective vevs $v_L$ and $v_R$. An additional bi-doublet $\phi\;(1,2,2)(0)$ scalar with vevs $v_{1,2}$ is  needed to provide mass to the fermions. Since both $v_1$ and $v_2$ are required to be non-zero for fermion mass generation, the  $W_L$ and $W_R$ gauge bosons of $SU(2)_L$ and $SU(2)_R$ will mix with each other. In terms of the mass eigenstates $W,Z$ and $W^\prime,Z^\prime$, the mixing can be parametrized as
\begin{eqnarray}
W_L = \cos\xi_W W - \sin \xi_W W^\prime && W_R = \sin\xi_W W +\cos\xi_W W^\prime,\nonumber\\
Z_L = \cos\xi_Z Z - \sin \xi_Z Z^\prime && Z_R = \sin\xi_Z Z +\cos\xi_Z Z^\prime
\end{eqnarray}
The model can provide potentially large contributions to amplitudes that involve third generation fermions when $g_R>>g_L$, where the new gauge bosons interact very weakly with the first two generations. The dominant gauge boson-fermion interactions in this limit are given by
\begin{eqnarray}
{\cal L}_W&=& 
-\frac{g_R}{ \sqrt{2}}
\bar U_{R} \gamma^\mu V_{R} D_{R}
(\sin\xi_W W^{+}_\mu + \cos\xi_W W^{'+}_{\mu}) ~+~{\rm h.~c.},\nonumber \\
{\cal L}_Z&=& \frac{g_L}{ 2} \tan\theta_W \cot\theta_R (\sin\xi_Z Z_\mu + \cos\xi_Z Z^\prime_\mu)
 \left(\bar d_{R_i} V^{d*}_{Rbi} V^d_{Rbj}\gamma^\mu d_{R_j} - \bar u_{R_i} V^{u*}_{Rti} V^u_{Rtj}\gamma^\mu u_{R_j} 
\right)\nonumber\\
&+& \frac{g_R}{ 2} \left(\bar\tau \gamma_\mu P_R \tau - N_R \gamma_\mu P_R N_R \right)~Z^\prime_\mu
\label{cccoup}
\end{eqnarray}
where 
$U = (u,\;\;c,\;\;t)$, $D = (d,\;\;s,\;\;b)$, $V_{KM}$ is
the Kobayashi-Maskawa mixing matrix and $V_R \equiv (V_{Rij})=(V^{u*}_{Rti}V^{d}_{Rbj})$ with $V^{u,d}_{Rij}$ the unitary matrices
which rotate the right handed quarks $u_{Ri}$ and $d_{Ri}$ from the weak  to the mass eigenstate basis. 

\subsection{Constraints}

This model has been studied at length before and here we summarize known constraints with appropriate numerical updates.

\begin{itemize}

\item{Perturbative unitarity.} To single out the third generation we require $g_R >> g_L$ and use perturbative unitarity near the electroweak scale to set the upper bound.  In the large $g_R$ limit\footnote{For a complete description of the model see  \cite{He:2002ha,He:2003qv}.} we find  
\begin{eqnarray}
\cot^2\theta_R=\left(\frac{g_R}{g_L}\cot\theta_W\right)^2-1&\implies g_L\tan\theta_W\cot\theta_R\approx g_R.
\end{eqnarray}
the two couplings appearing in Eq.~\ref{cccoup} are the same, and writing the perturbative unitarity condition as $g_R^2/2 \lesssim 4\pi$ results in 
$g_R \lesssim 5$. In order to compare with recent literature we will define the perturbative unitarity limit below in terms of the maximum resonance width, and this results in a similar number.

\item{LEP constraints.}  The most interesting limit from $e^+e^-\to \tau^+\tau^-$ and $e^+e^-\to b\bar b$ at LEP can be summarized by the relation \cite{He:2003qv}
\begin{eqnarray}
g_R M_{Z}\lesssim g_L M_{Z^\prime}.
\end{eqnarray}

\item{Gauge boson mixing.} For the model to be phenomenologically viable, both $\xi_W$ (from $B\to X_s\gamma$ \cite{He:2002ha}) and $\xi_Z$ (from $Z\to \tau\tau$ \cite{He:2003qv}) are required to be very small
\begin{eqnarray}
-0.0014\lesssim \frac{g_R}{g_L}\xi_W\lesssim 0.0018, && \left|\frac{g_R}{g_L}\xi_Z\right|\lesssim 3\times 10^{-3}
\end{eqnarray}
and we will set them to zero for the remainder of this study. 

\item{Right handed quark mixing angles.} The mixing angles in $V^{u,d}_R$ control the size of tree-level flavor changing neutral currents of the $Z^\prime$ and are also severely constrained. The complete set of constarints can be found in \cite{He:2004it,He:2006bk}, but the most relevant for this study are obtained by combining $B_s$ and $B_d$ mixing to yield
\begin{eqnarray}
|V^{d\star}_{Rbs}V^d_{Rbd}|\lesssim 6.7\times 10^{-7}
\label{cbmix}
\end{eqnarray}
This constraint (and others found in \cite{He:2004it,He:2006bk}) can be satisfied in a simple manner with the ansatz described in Ref.~\cite{He:2009ie}, which in turn yields predictions for $V^u_{Rtj}$. In particular, for this study we satisfy Eq.~\ref{cbmix} as $V^{d}_{Rbj}=\delta_{bj}$ and the ansatz then results in $V^{u}_{Rtt}\sim 1$, $V^{u}_{Rtc}\sim V_{cb}$, $V^{u}_{Rtu}\sim V_{ub}$. Importantly, it also results in  $V^{u}_{Rtu}V^d_{Rbd}\lesssim {\rm few}~\times 10^{-7}$, which implies that $W^\prime$ production from light-quark annihilation is completely negligible at LHC.

\end{itemize}

When we combine the above results with $g_R>>g_L$ the couplings in Eq.~\ref{cccoup} reduce to 
\begin{eqnarray}
{\cal L}_{W^\prime}&=& -\frac{g_R}{ \sqrt{2}}\left(
\bar t_{R} \gamma^\mu b_{R}+V_{cb}^\star~\bar c_{R} \gamma^\mu b_{R}+N_R \gamma_\mu P_R \tau\ \right)~
 W^{'+}_{\mu} ~+~{\rm h.~c.},\nonumber \\
{\cal L}_{Z^\prime}&=& \frac{g_R}{ 2} 
 \left(\bar b_{R} \gamma^\mu b_{R} - \bar t_{R} \gamma^\mu t_{R} - V_{cb}\bar t_{R} \gamma^\mu c_{R}- V_{cb}^\star\bar c_{R} \gamma^\mu t_{R}\right)~Z^\prime_\mu \nonumber \\
&+& \frac{g_R}{ 2} \left(\bar\tau \gamma_\mu P_R \tau - N_R \gamma_\mu P_R N_R \right)~Z^\prime_\mu
\label{cccoupsimp}
\end{eqnarray}
Eq.~\ref{cccoupsimp} defines the simplified model used in this study.

\subsection{Resonance width and branching ratios}

A recent model independent analysis of LHC constraints on a $W^\prime$ with dominant couplings to third generation fermions \cite{Greljo:2018tzh} serves as motivation for this study. In order to compare with that paper, or to cast our study in a more model independent way, we can trade the couplings of the $W^\prime$ in the limit discussed above for the resonance width, this results in
\begin{eqnarray}
\frac{\Gamma_{W^\prime}}{M_{W^\prime}}&=&(1+N_c)\frac{g_R^2}{48\pi}.
\label{wwidth}
\end{eqnarray}
The second term, proportional to $N_c$, corresponds to the top-bottom channel for which we assume $M_{W^\prime}>>M_t$ and the first term to the  $\tau N_R$ channel where we assume that the sterile neutrino is light, $M_{W^\prime}>>M_{N_R}$. 
Defining perturbative unitarity as Ref.~\cite{Greljo:2018tzh}  does, requiring the width to mass ratio be smaller than 1/2, results in 
\begin{eqnarray}
g_R \lesssim 4.34 {\rm ~or~}\left(\frac{g_R}{g_L}\right)\lesssim 6.7
\label{unitarity}
\end{eqnarray}
which is within the range that has been previously discussed. In this large $g_R$ limit, one also finds that
\begin{eqnarray}
{\cal B}(W^\prime \to \tau N_R)\approx 25\%\frac{4P_{W^\prime}}{3 + P_{W^\prime}}.
\label{brwp}
\end{eqnarray}
where we have now included a phase space factor 
\begin{eqnarray}
P_{W^\prime} = \left(1-\frac{M^2_N}{M^2_{W^\prime}}\right) \left(1-\frac{M^2_N}{2M^2_{W^\prime}} - \frac{ M^4_N}{2M^4_{W^\prime}}\right)
\end{eqnarray}
to correct for the mass of the sterile neutrino when it is large but still below $M_{W^\prime}$.

It is also useful to express the $Z^\prime$ gauge coupling in terms of its width. In the limit where $g_R$ is large, the dominant contributions will be from decays into third generation fermions plus the new right-handed neutrino,  resulting in
\begin{eqnarray}
\frac{\Gamma_{Z^\prime}}{M_{Z^\prime}}=\left(1+N_c\right)\frac{g_R^2}{48\pi}.
\label{zwidthgr}
\end{eqnarray}
The second term arises from decays into bottom and top-pairs, again assuming $M_{Z^\prime}>>2M_t$, and the first term from decays into tau-lepton pairs and  $N_R$ pairs assumed to be light. Note that, with the assumptions listed above, the $Z^\prime$ and $W^\prime$ widths are the same. If the sterile neutrino mass is not negligible but still below half the $Z^\prime$ mass, the $Z' \to N_R \bar N_R$ mode would be suppressed by the phase factor
\begin{eqnarray}
P_{Z^\prime} = \left(1-\frac{M^2_N}{M^2_{Z^\prime}}\right) \sqrt{1-\frac{4M^2_N}{M^2_{Z^\prime}}}
\end{eqnarray}
resulting in
\begin{eqnarray}
{\cal B}(Z^\prime \to \tau\tau)\approx \frac{1}{7 + P_{Z^\prime}}.
\end{eqnarray}

\begin{figure}[h!]
\centering{\includegraphics[width=.48\textwidth]{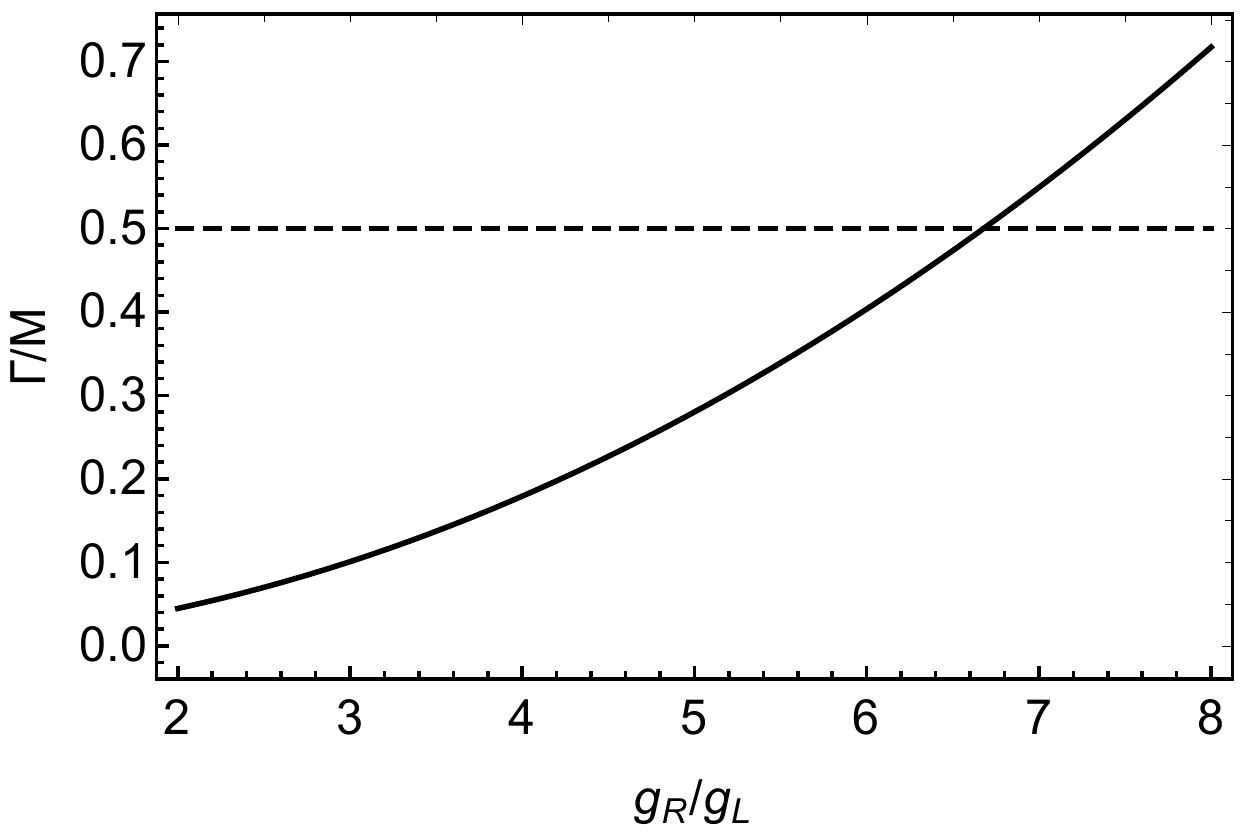}}
\caption{$\Gamma/M$ for both $W^\prime$ and $Z^\prime$ as a function of $g_R/g_L$. The horizontal dashed line is adopted as the boundary for perturbative unitarity.}
\label{res-width}
\end{figure}

The region of parameter space that is of interest for this study has $g_R >> g_L$ and from Eq.~\ref{wwidth} and Eq.~\ref{zwidthgr} we see this corresponds to relatively fat resonances as can be seen in Figure~\ref{res-width}. Assuming that the resonances are much larger than both the top-quark and the sterile neutrino masses, the figure shows the ratio $\Gamma/M$ for both $W^\prime$ and $Z^\prime$ as a function of $g_R/g_L$. The dashed horizontal line marks the limit we use for perturbative unitarity.

\section{Signatures at the LHC}

\begin{figure}[h!]
\centering{\includegraphics[width=.45\textwidth]{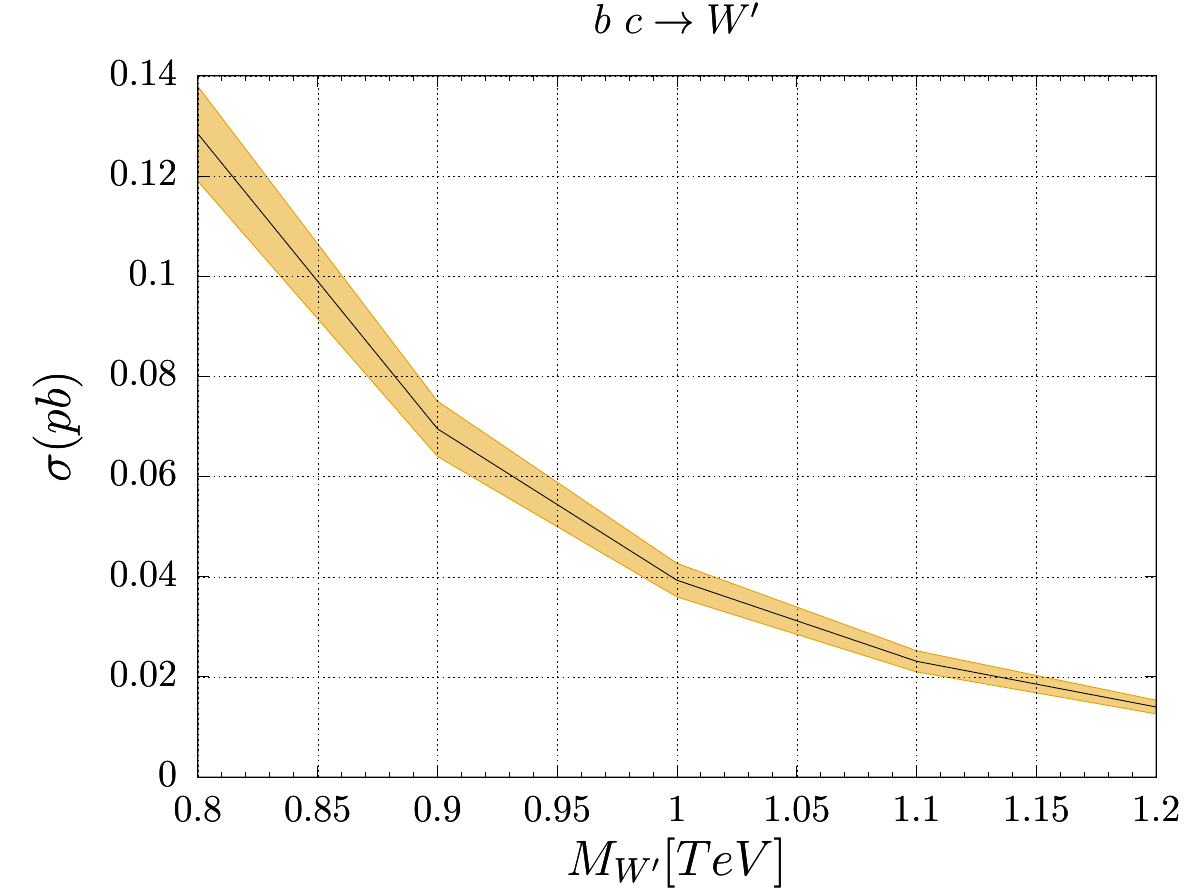}~~~\includegraphics[width=.45\textwidth]{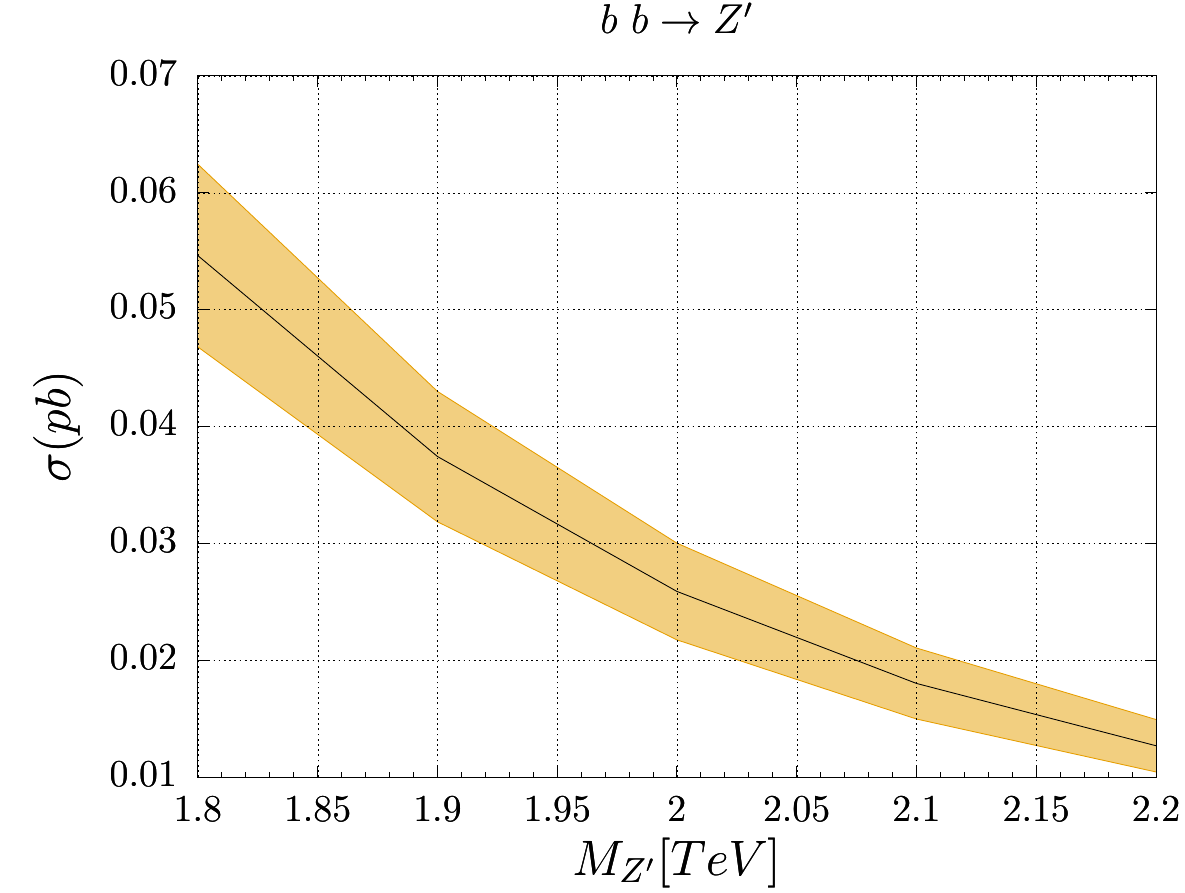}}
\caption{Production cross-section $1\sigma$ uncertainty bands for  $pp(b\bar{c})\to W^\prime$ (left) and  $pp(b\bar{b})\to Z^\prime$ (right) obtained for $g_R=3.96$ and $g_R=3$ respectively. The band only reflects the uncertainty in the parton distribution functions. The couplings and mass ranges have been chosen to illustrate the more interesting regions found in the next subsections.}
\label{pdfun}
\end{figure}

For our numerical study to recast the LHC constraints we implement the Lagrangian of  Eq.(\ref{cccoupsimp}) in FEYNRULES \cite{Christensen:2008py,Degrande:2011ua} to generate a Universal Feynrules Output (UFO) file, and then feed this UFO file into MG5\_aMC@NLO \cite{Alwall:2014hca}. We have used the  PDF4LHC15\_nlo\_mc set of parton distribution functions including for the $b$-quark. A few generic remarks apply to the level of uncertainty in this numerical study. 
 The largest systematic uncertainty in the parton-level simulation is the PDF uncertainty as already argued in Ref.~\cite{Greljo:2018ogz}.  We illustrate this in Figure~\ref{pdfun}, where we show the $1\sigma$ uncertainty band obtained by using PDF errorset replicas. The figure illustrates production cross-sections $pp(b\bar{c})\to W^\prime$ (left) and  $pp(b\bar{b})\to Z^\prime$ (right) for the mass regions of interest to the constraints we obtain later. For $W^\prime$ production with $g_R=3.96$ the uncertainty for the mass range shown ranges from 7.4\% to 9.8\%, whereas for $Z^\prime$ production with $g_R=3$ it ranges from 14.3\% to 17.7\%. In addition, 
the flavor physics motivations for this study suggest looking at coupling constant values near their perturbative bound. This implies that the model tree-level calculations also have a large uncertainty. In addition, the production of $W^\prime$ depends on a subleading unknown parameter in our model, $V^{u}_{Rtc}$. We have argued before that this is expected to be of order and $V^{u}_{Rtc}\sim V_{cb}$ and used this for our estimates. But it is important to keep in mind that this parameter can also be treated as a free parameter that can change the  $W^\prime$ production cross-section without affecting its width significantly.

\subsection{$W^\prime$}

We begin with the right-handed $W^\prime$ with couplings as in Eq.~\ref{cccoupsimp}.  Single production of this $W^\prime$ at the LHC is then dominated by $c\bar{b}~(b\bar{c})$ annihilation as the couplings to $u,d$ quarks are zero in this approximation. The dominant decays would be into $\tau N_R$ if kinematically allowed, and into $t\bar{b} (b\bar{t})$. Searches for $W^\prime$ bosons in both of these channels have been carried out: by CMS in the $\tau_h+p_T^{\rm miss}$ final state  with 35.9 fb$^{-1}$ at 13~TeV \cite{Sirunyan:2018lbg}, and by ATLAS in the  $t\bar{b} (b\bar{t})$ final state with 36.1 fb$^{-1}$ at 13~TeV \cite{Aaboud:2018jux}.

The parameter region of interest corresponds to fat resonances (see Figure~\ref{res-width}) and consequently the narrow width approximation is not expected to be reliable. We compute the cross-section $\sigma(pp\to \tau N_R)$ with {\tt MadGraph}, allowing contributions well outside the resonance width but working with an energy independent width. The calculation defined in this way corresponds purely to new physics and the cross-section scales approximately as $g_R^2$. The use of an energy independent width for fat spin zero resonances is known to overestimate the cross-section in the vicinity of the resonance  \cite{Valencia:1992ix,Seymour:1995np}. We multiply $\sigma(pp\to \tau N_R)$ by the branching ratio for hadronic tau decay, approximately 0.65, in order to compare our results with  Figure~5 of  \cite{Sirunyan:2018lbg} (CMS). We expect this calculation to overestimate the  cross-section due to the use of an energy independent width and to our neglect of detector effects, in this sense our limits will be conservative. The use of the specific final state with the sterile neutrino removes interference terms with the SM that could occur in more general models.

\begin{figure}[h!]
\centering{\includegraphics[width=.45\textwidth]{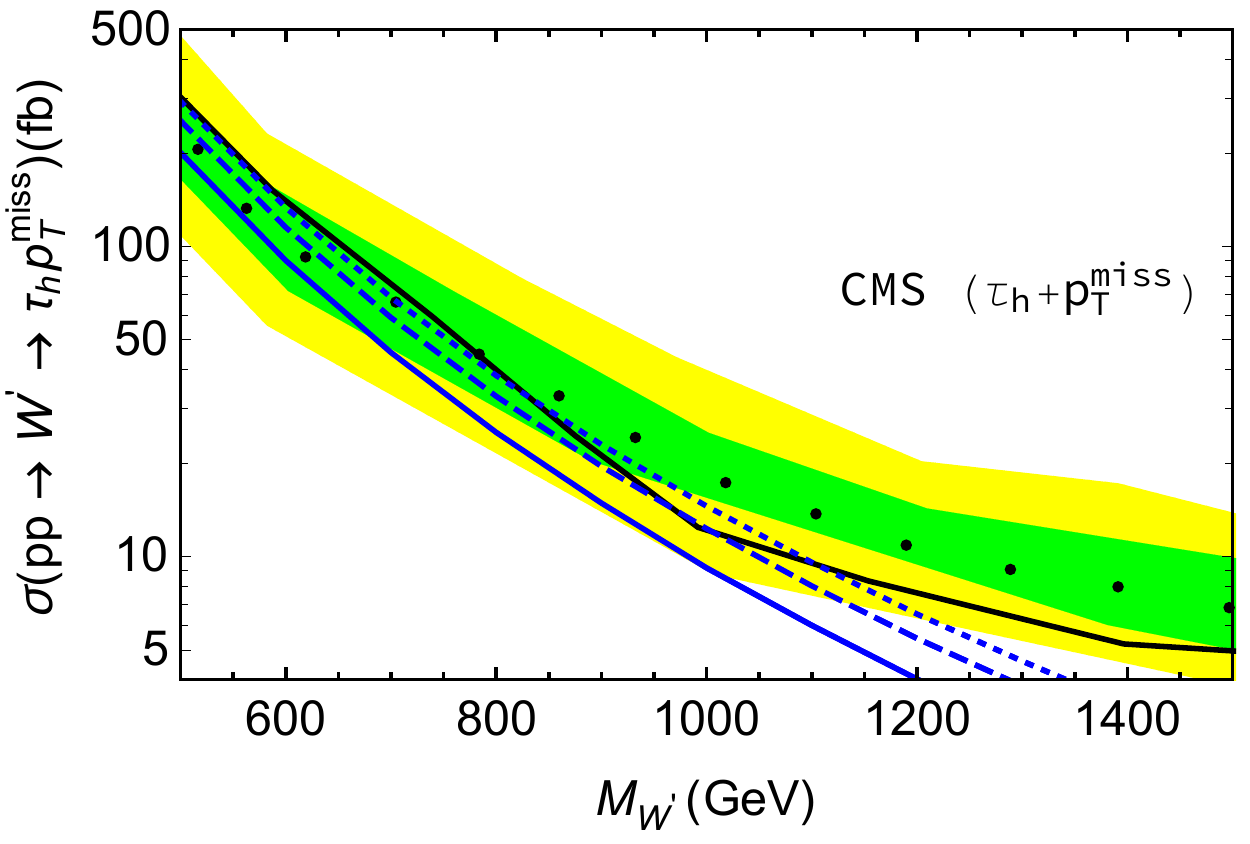}~~~\includegraphics[width=.45\textwidth]{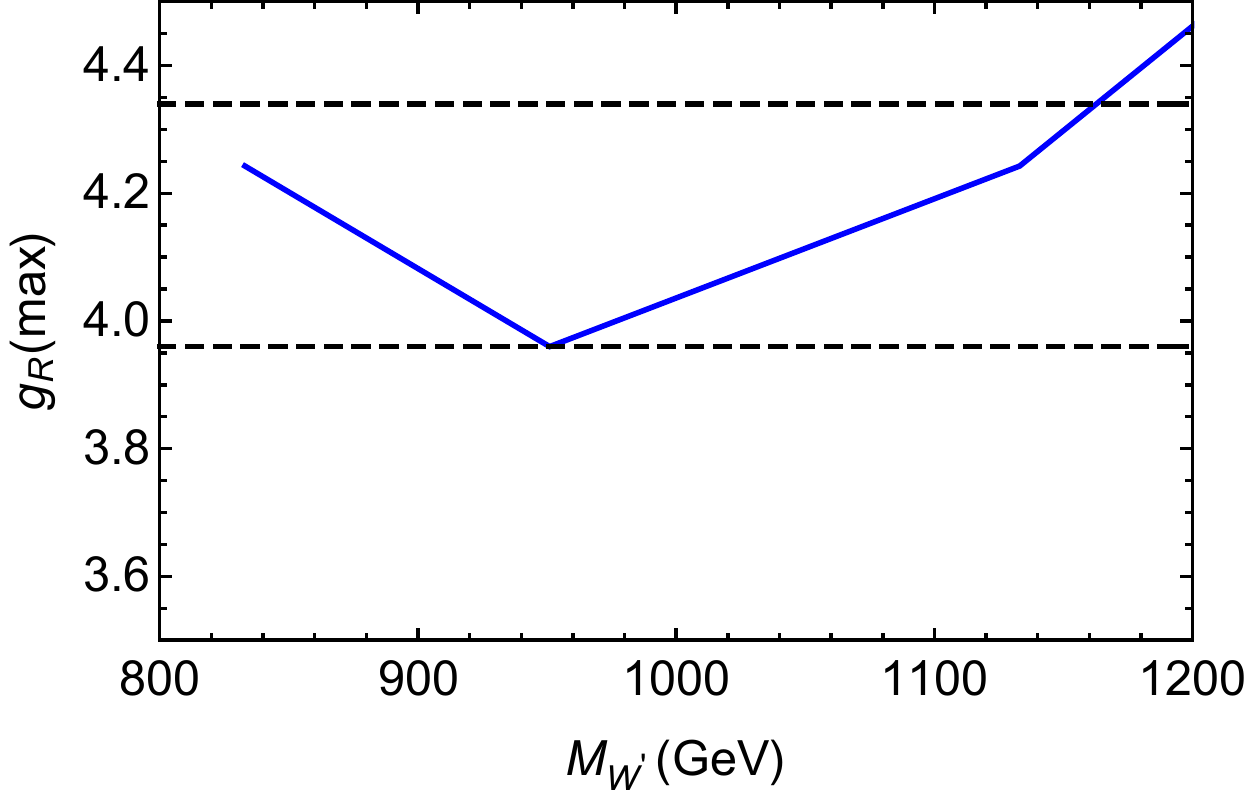}}
\caption{Left panel: $\sigma(pp\to W^{\prime\pm}\to \tau\nu)$ for $g_R=3.53,3.96,4.24$ ($g_R/g_L=5.4,6.1,6.5$) solid, dashed and dotted blue lines respectively, superimposed on the CMS results: 95\% CL observed (solid), expected (dashed), 1$\sigma$ (green) and 2$\sigma$ (yellow) (Figure~5 of \cite{Sirunyan:2018lbg}). The right panel shows the maximum value of $g_R$ allowed by the CMS observed 95\% CL bound for a given $W^\prime$ mass. The horizontal dashed lines mark the strongest constraint and the perturbative unitarity limit on $g_R$.}
\label{f:wplimit}
\end{figure}

Our results for $\sigma(pp\to W^\prime \to \tau_h p_T^{\rm miss})$ are presented in Figure~\ref{f:wplimit}. The left panel illustrates our  cross-sections for values $g_R/g_L=5.4,6.1,6.5$  as the solid, dashed and dotted blue lines. These model calculations are superimposed on the CMS limit: 95\%cl observed (solid), expected (dashed), 1$\sigma$ (green) and 2$\sigma$ (yellow) taken from Figure~5 of  \cite{Sirunyan:2018lbg}. The result exhibits an interesting behaviour: the largest value of $g_R/g_L$ shown corresponds to $g_R=4.24$ and is near the perturbative unitarity limit.  The cross-section in this case  lies within the expected $1\sigma$ exclusion for any value of $M_{W^\prime}$ below about 1~TeV. However, it lies {\it above}  the CMS observed 95\% CL bound only for the range $826 \lesssim M_{W^\prime}\lesssim 1100$~GeV. Taken at face value this means that both light and heavy resonances are allowed at the 95\% CL. The dashed curve corresponds to $g_R=3.96$ and constitutes a limiting case: it barely touches the CMS observed 95\% CL bound at $M_{W^\prime}\approx 990$~GeV. Values of $g_R\lesssim 3.96$ are therefore allowed for any value of $M_{W^\prime}$. The center panel shows the allowed region in $M_{W^\prime}-g_R$ parameter space, the area below the blue line. The boundary, labelled $g_R({\rm max})$, is obtained from the intersection between the $\sigma(pp\to W^\prime\to \tau_h\nu)$ curve in the model and the CMS observed 95\% CL bound. This line then represents the constraints from this process for the narrow window between $3.96\lesssim g_R\lesssim 4.34$ (where we reach the perturbative unitarity limit).  This perturbative unitarity limit, as well as $g_R=3.96$ (the lowest value that is constrained by the CMS data) are shown as horizontal dashed lines.  
The ATLAS results \cite{Aaboud:2018vgh} are currently less restrictive than CMS and do not change this picture.

The $W^\prime$ in the models considered in \cite{Sirunyan:2018lbg} is left-handed and narrow, whereas our $W^\prime$ is right-handed and fat and these differences could affect the analysis. In  particular, the right-handed $\tau$-lepton produced in our model, has different decay kinematics from its left-handed counterpart. To take this into account we turn to the model independent limit produced by CMS (Figure 8 of \cite{Sirunyan:2018lbg}). The CMS analysis assumes a certain shape for the $m_T$ distribution, where 
\begin{eqnarray}
m_T=\sqrt{2p_T^\tau p_T^{\rm miss}(1-\cos\Delta\phi(\vec{p}_T^\tau,\vec{ p}_T^{\rm miss})}.
\end{eqnarray}
To correct for a different $m_T$ distribution, one needs to compare the calculated $\sigma{\cal B}$ against the CMS result applying a correction factor $f_{m_T}(m_T^{\rm min})$, given by   the fraction of generated events with $m_T>m_T^{\rm min}$ as calculated in our model.  To calculate this factor we first process our events with Pythia 8 \cite{Sjostrand:2006za,Sjostrand:2007gs} which includes  fully modeled hadronic tau-lepton decays based in Tauola and Herwig++ \cite{Ilten:2012zb,Jadach:1993hs,Grellscheid:2007tt}. We then analyse the results with Rivet \cite{Buckley:2010ar} and at this stage we can apply the final cuts used by CMS: $p_T^\tau \geq 80$~GeV; $p_T^{\rm miss} > 200$~GeV and $\Delta\phi(\vec{p}_T^\tau,\vec{ p}_T^{\rm miss})> 2.4$~rad.  

\begin{figure}[h!]
\centering{\includegraphics[width=.45\textwidth]{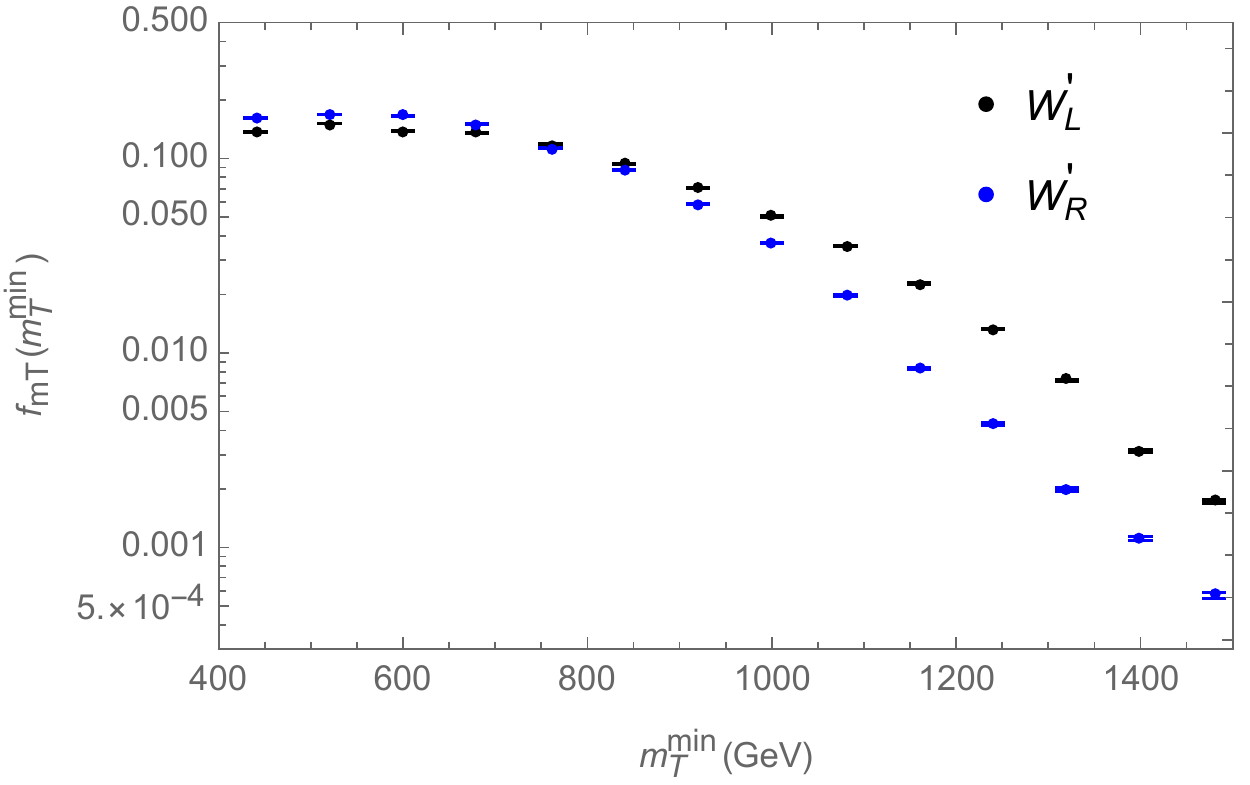}~~~\includegraphics[width=.45\textwidth]{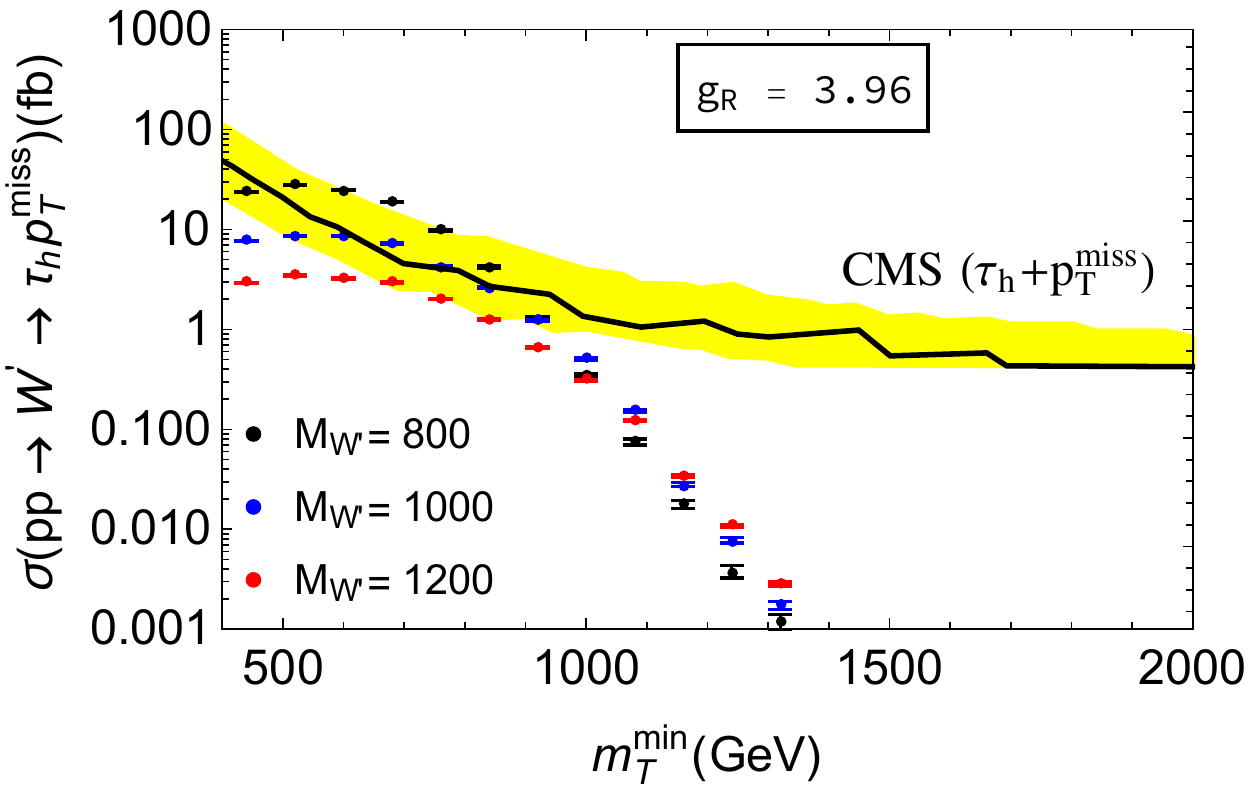}}
 \caption{Left panel:  $f_{m_T}(m_T^{\rm min})$ for a left and right-handed $W^\prime$ of mass 1~TeV and width corresponding to $g_R=3.96$. Right panel: comparison of $\sigma(pp\to W^{\prime\pm}\to \tau\nu)$ for $g_R=3.96$ and $M_{W^\prime}$= 800~GeV (black), 1 TeV (blue) and 1.2~TeV (red) with the model independent limits of CMS: 95\% CL observed (solid),  2$\sigma$ (yellow) from Figure 8 of \cite{Sirunyan:2018lbg}.}
 \label{f:milimit}
\end{figure}

Our results are shown in Figure ~\ref{f:milimit}. In the left panel we compare $f_{m_T}(m_T^{\rm min})$ for a left and right-handed $W^\prime$ of mass 1~TeV and width corresponding to $g_R=3.96$ in our model, with statistical uncertainties on a sample of 100k events. For the left-handed case we use a non-SM neutrino to insure that there is no interference with the SM. In the right panel we compare our model results for three different $W^\prime$ masses to the CMS model independent limit. From this figure, we conclude that CMS restricts $M_{W^\prime}\lesssim 1$~TeV for $g_R=3.96$ which compares with $M_{W^\prime}\lesssim 950$~GeV shown in Figure \ref{f:wplimit}. Two comments are in order: first, our resuts using the true $\tau$ momentum produce limits that are a bit conservative (5\% in this specific case); and second, the different limits fall within the PDF uncertainties as shown in Figure~\ref{f:res}.

\begin{figure}[h!]
\centering{\includegraphics[width=.48\textwidth]{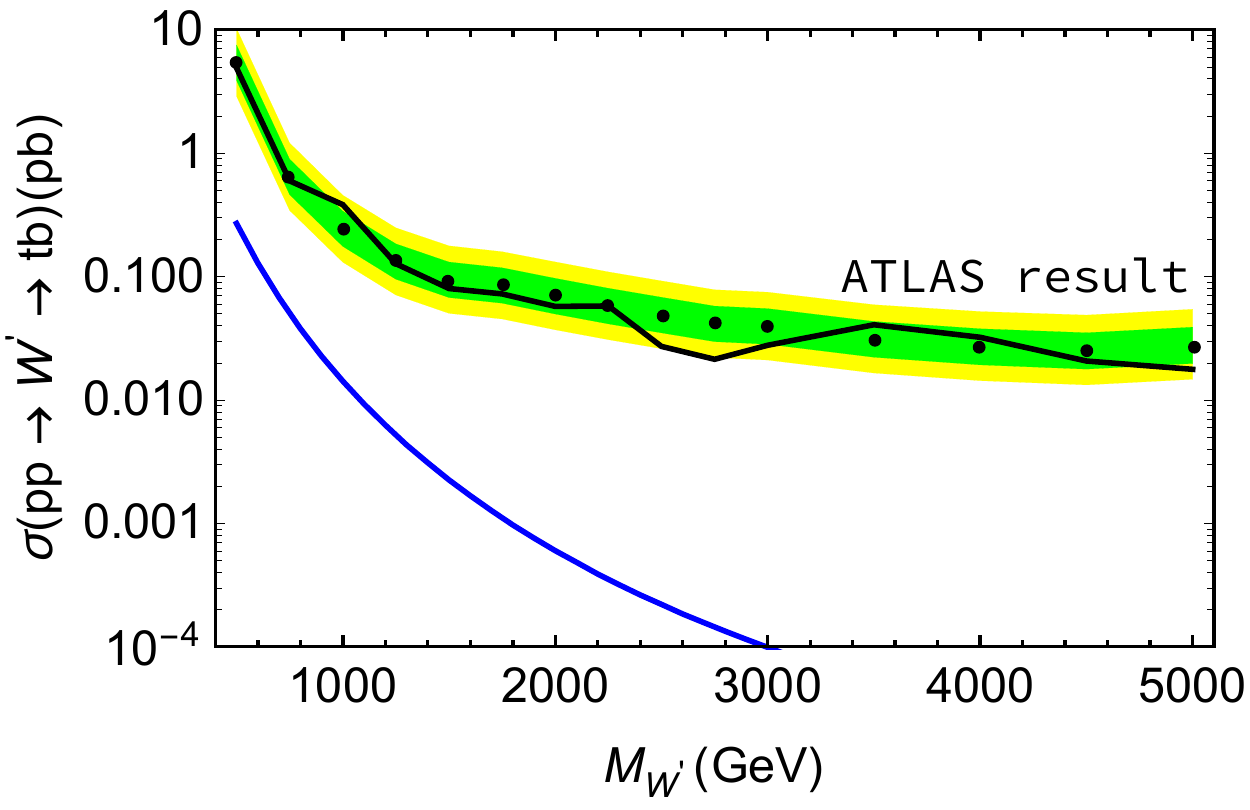}}
\caption{$\sigma(pp\to W^{\prime\pm}\to t\bar{b} )$  for $g_R=4.24$ shown as a blue line, superimposed on the ATLAS limit: 95\% CL observed (solid), expected (dashed), 1$\sigma$ (green) and 2$\sigma$ (yellow) from Figure~8 of \cite{Aaboud:2018jux}). The data does not yet constrain the model.}
\label{f:wplimittb}
\end{figure}

The other important decay channel in the model is $W^{\prime+}\to t\bar{b}$, which becomes the dominant one for a heavy sterile neutrino. It can be constrained in principle from the cross-section $\sigma(pp\to W^{\prime\pm}\to t\bar{b} )$. The current ATLAS result from 36.1 fb$^{-1}$ at $\sqrt{S}=13$~TeV (combination of semileptonic and hadronic searches) does  not yet  constrain this model  as illustrated in Figure~\ref{f:wplimittb} where we show the model prediction for $g_R=4.24$ (near its perturbative unitarity bound) in blue.

\subsection{$Z^\prime$}

The production mechanism for the $Z^\prime$ is less model dependent than that of the $W^\prime$ as the leading flavour diagonal couplings to fermions do not depend on unknown mixing angles. In the limit of Eq.~\ref{cccoupsimp} the dominant production at LHC is  initiated by $b\bar{b}$ annihilation. The main final states of interest are $t\bar{t}$ and $\tau^+\tau^-$, a $b\bar{b}$ final state is also possible but its study is more difficult due to QCD backgrounds \footnote{Hadron colliders constraints on this type of $Z^\prime$ produced from light-quark annihilation were studied in Refs.~\cite{Han:2003pu,Han:2004zh}.}

\begin{figure}[h!]
\centering{\includegraphics[width=.48\textwidth]{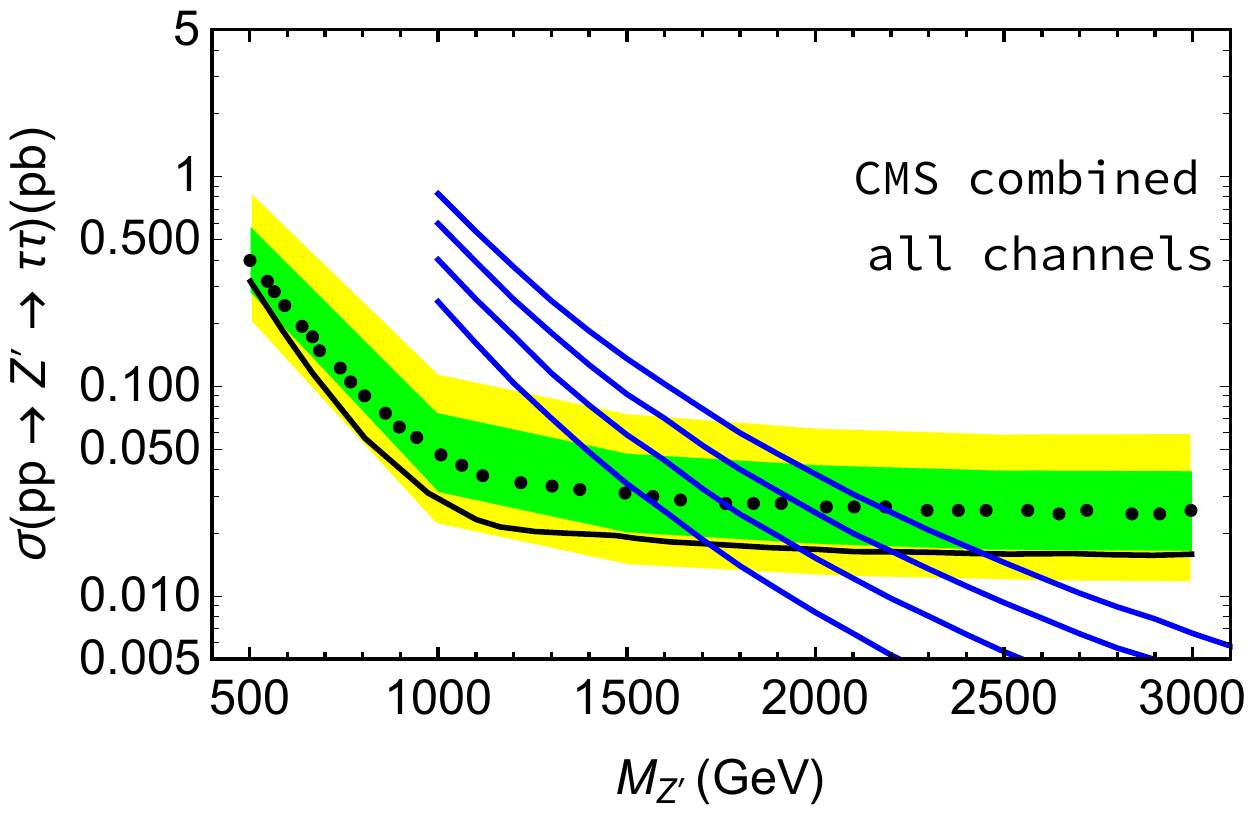}~~~\includegraphics[width=.48\textwidth]{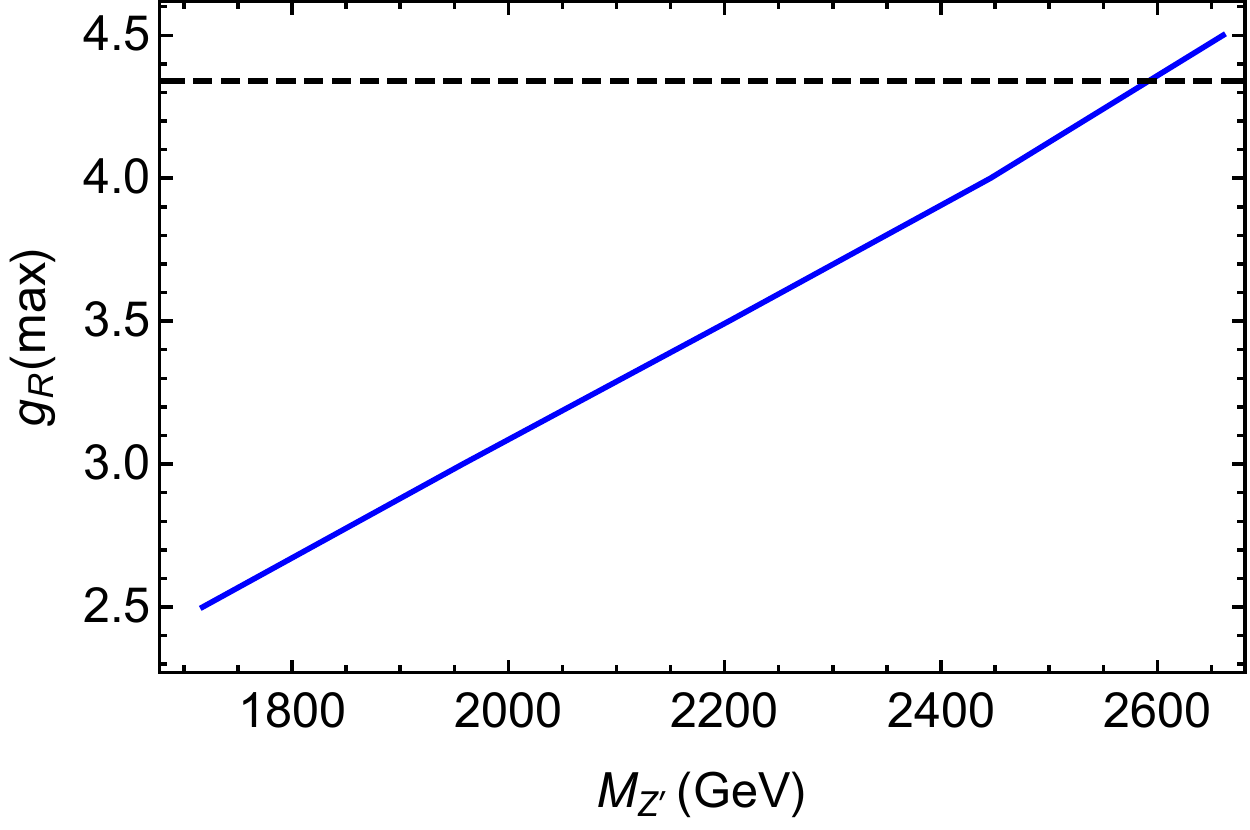}}
\caption{Left panel: we reproduce the 95\% CL observed (solid), expected (dotted), 1$\sigma$ (green) and 2$\sigma$ (yellow) from CMS Figure~2e of \cite{Khachatryan:2016qkc} for the process $\sigma(pp\to Z^\prime\to\tau^+\tau^-)$ and superimpose the cross-sections predicted in our model for representative values $g_R=2.5,3,3.5,4$ shown in blue (with increasing values of $g_R$ from the leftmost curve) that follow from Eq.~\ref{cccoupsimp}. Right panel: upper limit on $g_R$ as a function of $M_{Z^\prime}$ as read from the left panel.}
\label{f:zpcms}
\end{figure}

In Figure~\ref{f:zpcms} we compare our $Z^\prime$ production from $b\bar{b}$ annihilation followed by decay into di-tau pairs in all tau decay channels  with the CMS result of Ref.~\cite{Khachatryan:2016qkc}.  On the left panel we superimpose the cross-sections $\sigma(pp\to Z^\prime\to\tau^+\tau^-)$ for representative values $g_R=2.5,3,3.5,4$ shown in blue (with increasing values of $g_R$ from the leftmost curve) onto  Figure~2e of \cite{Khachatryan:2016qkc}. We treat the fat resonance in the same manner as before: use an energy independent width and compute $\sigma(pp\to \tau^+\tau^-)$ allowing for contributions away from the resonance but not including interference with the SM. 
The right panel shows the maximum value $g_R$ can take for a given $M_{Z^\prime}$ as determined by requiring the model cross-section to be below the 95\% CL limit observed by CMS, and the dashed line marks the perturbative unitarity limit on $g_R$.

\begin{figure}[h!]
\centering{\includegraphics[width=.48\textwidth]{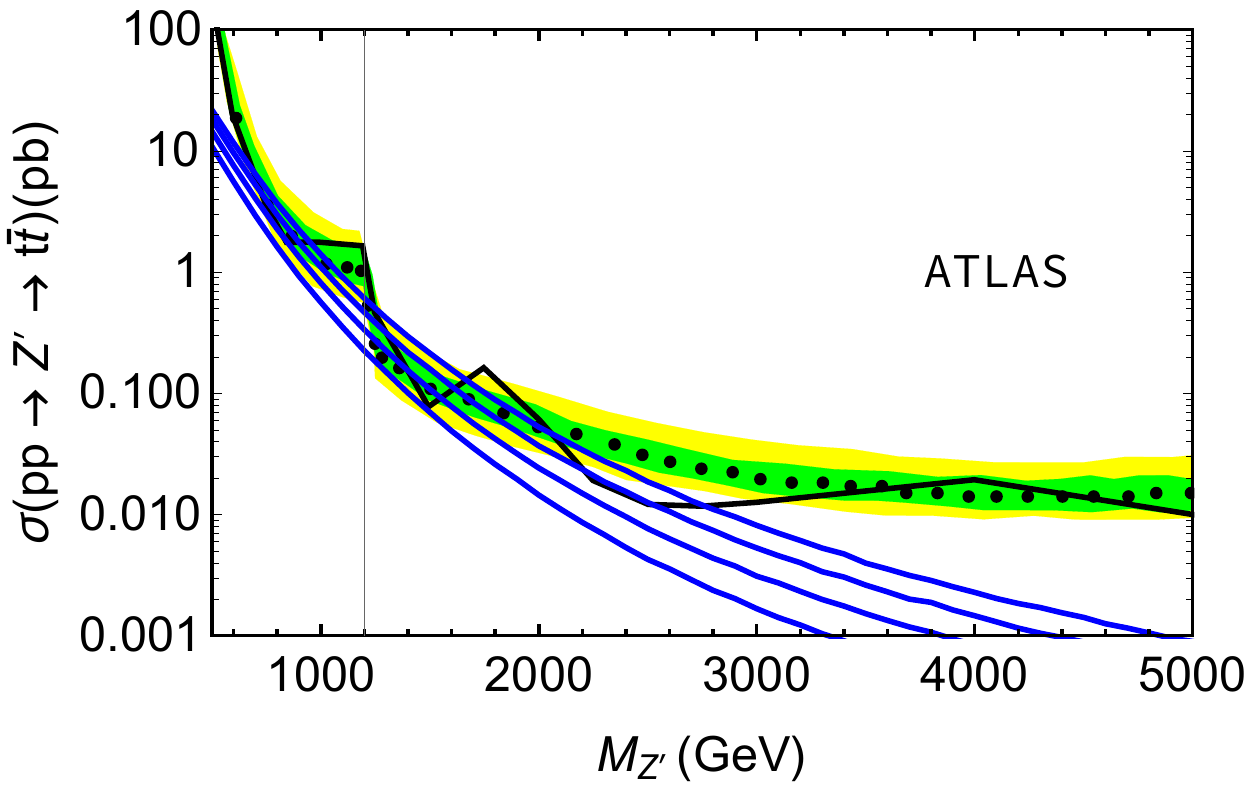}~~~\includegraphics[width=.48\textwidth]{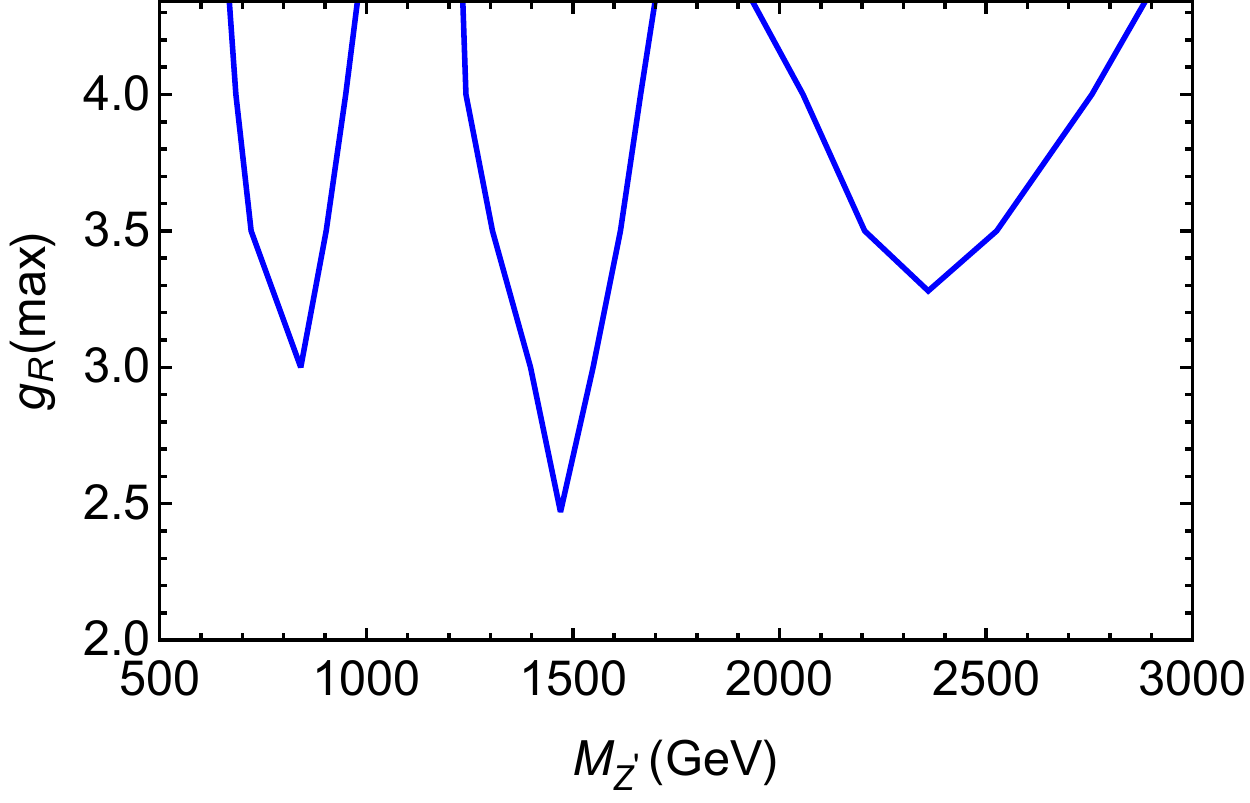}}
\caption{Left panel: 95\% CL observed (solid), expected (dotted), 1$\sigma$ (green) and 2$\sigma$ (yellow) limits on $\sigma(pp\to Z^\prime\to t \bar{t})$ from ATLAS Figure~9 of \cite{Aaboud:2019roo} on which we superimpose the cross-sections predicted in our model for representative values $g_R=2.5,3,3.5,4$ (shown in blue with increasing values of $g_R$ starting from the lowermost curve). Right panel: upper limit on $g_R$ as a function of $M_{Z^\prime}$ as read from the left panel.}
\label{f:zpatlas}
\end{figure}

In Figure~\ref{f:zpatlas} we compare $t\bar{t}$ production  from $b\bar{b}$ annihilation through an intermediate $Z^\prime$ with the ATLAS result shown in Figure~9 of \cite{Aaboud:2019roo}. The ATLAS figure is split into a resolved and boosted analysis by the vertical line at 1200~GeV. The shape of the observed limit makes it difficult to interpret the constraints this result imposes on our model. Taken at face value, $g_R$ cannot exceed the value for which the prediction intersects the 95\%CL observed ATLAS limit (solid black line in the left panel) and this is depicted on the right panel.
The figure suggests that masses below about $1500$~GeV are excluded for the parameter region with $g_R >> g_L$, although strictly speaking  there remains a window near 1~TeV with no meaningful constraints. For masses above 2.2~TeV $g_R\lesssim 3$ is allowed and for masses above 2.8~TeV all values of $g_R\lesssim 4.34$ (in the perturbative region) are allowed. 

\subsection{Four third generation fermion production at LHC}

An ideal way to test this type of model is to search for final states with four third generation fermions in the final state. The enhanced signal for these channels would occur through resonance pair production or through resonance production in association with a heavy quark pair. Some of these channels have already been identified as promising for this model in Ref.~\cite{Han:2003pu,Han:2004zh,Hayreter:2017wra}, but at present there is no constraining data. 

\begin{figure}[h!]
\centering{\includegraphics[width=.48\textwidth]{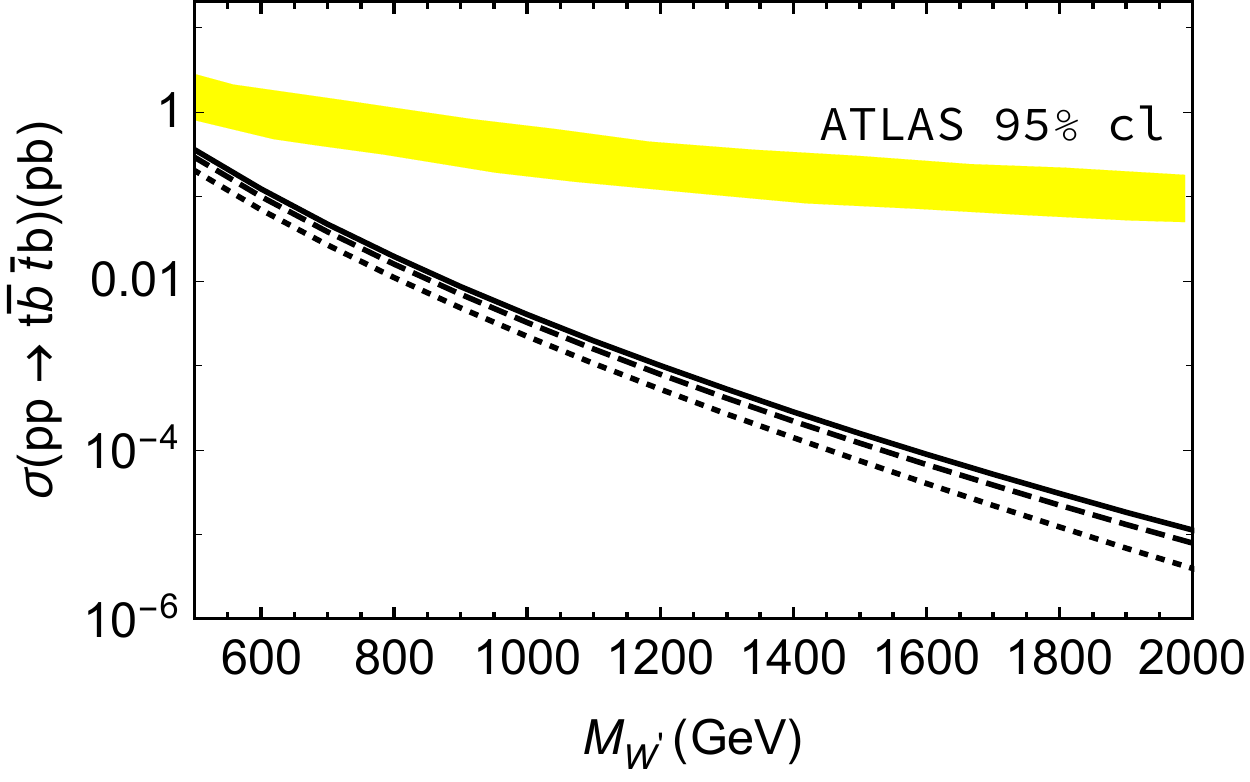}~~~\includegraphics[width=.48\textwidth]{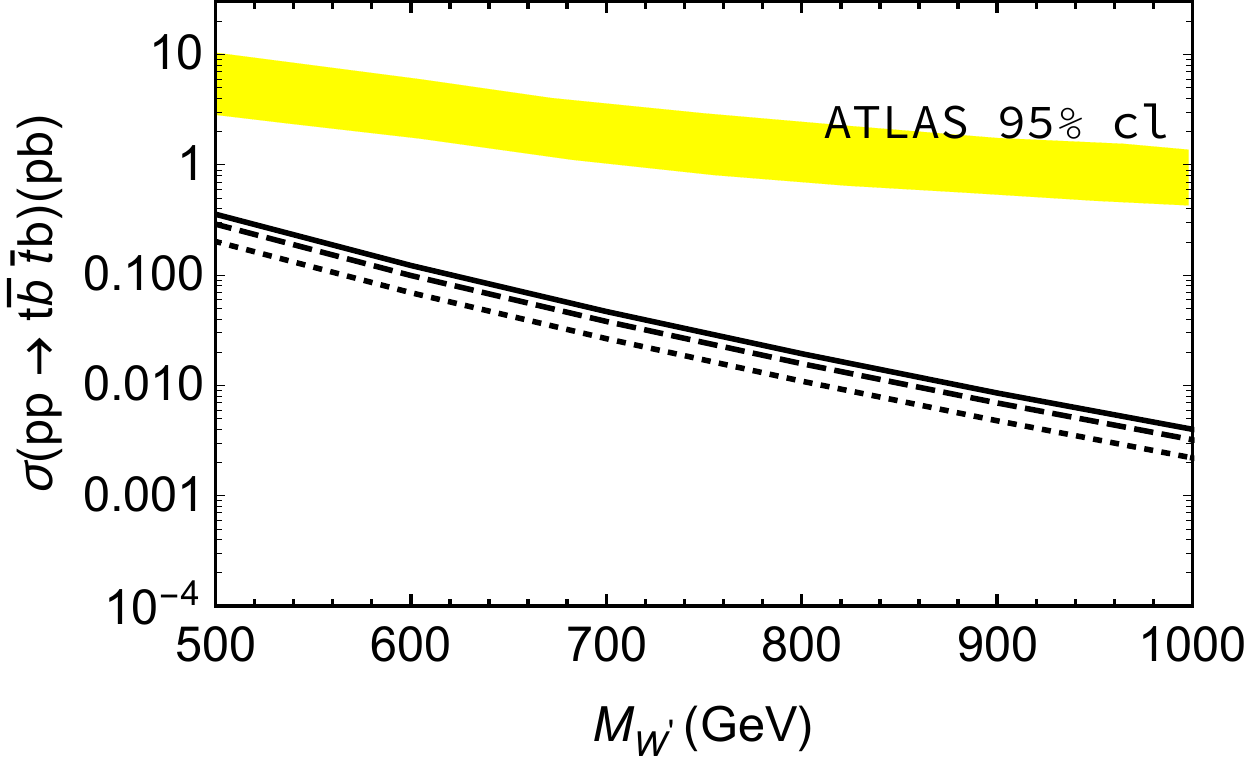}}
\caption{Model predictions for $pp\to W^{\prime +}W^{\prime -} \to t\bar{b}\bar{t}b$ for $g_R=3.53,~3.96,~4.24$ shown respectively as dotted, dashed and solid black lines. Left panel: 95\% CL expected limits on $pp\to t\bar{b}H^+\times BR(H^+\to t\bar{b})$ from ATLAS Figure~23 of \cite{ATLAS:2016btu}. 
 Right panel: model predictions shown with 95\% CL expected limits on $pp\to b\bar{b}H\times BR(H\to t\bar{t})$ from ATLAS Figure~21 of \cite{ATLAS:2016btu}.}
\label{f:wwp}
\end{figure}

A particularly interesting channel for this study is $W^{\prime +}W^{\prime -}$ production which has a model independent component through the electromagnetic coupling. We have computed the cross-section for $pp\to W^{\prime +}W^{\prime -} \to t \bar b \bar t b$ for illustration, as this final state has the most constraining results at present. The photon intermediate state is not the only contribution, in the limit of Eq.~\ref{cccoupsimp} there are also $b\bar b$ and $c \bar c$ initiated t-channel diagrams contributing. For the parameter range discussed so far, $3.0\lesssim g_R\lesssim 4.3$,  the photon contribution is largest for the smaller values of $g_R$ and the largest values of $M_{W^\prime}$ and does not exceed 34\%. This means that the pair production cross-section is also model dependent. In Figure~\ref{f:wwp} we compare the model results for a range of parameters to existing bounds on $pp\to t \bar b \bar t b$ from two ATLAS searches. The dotted, dashed, solid black lines are the model predictions for $g_R=3.53,~3.96,~4.24$ respectively and the yellow band on the left figure  is the 95\% cl expected limit from 13.2~fb$^{-1}$ at $\sqrt{S}=13$~TeV collected by ATLAS for the search $pp\to t\bar{b}H^+\times BR(H^+\to t\bar{b})$ \cite{ATLAS:2016btu}. The yellow band on the right panel is for the same final state but from the study of $pp\to b\bar{b}H\times BR(H\to t\bar{t})$ \cite{ATLAS:2016btu}. 
The figure indicates that this comparison will have to await more sensitive collider searches conducted specifically for $pp\to W^{\prime +}W^{\prime -}$ kinematics to constrain the model.

\section{Flavour Physics consequences}

\subsection{$R(D)$ and $R(D^\star)$}

A  $W^\prime$  in conjunction with a light sterile neutrino has been proposed as a possible explanation of the 
$R(D)$ \cite{Lees:2012xj,Lees:2013uzd,Huschle:2015rga}  and $R(D^\star)$  \cite{Lees:2012xj,Lees:2013uzd,Huschle:2015rga,Sato:2016svk,Hirose:2016wfn,Aaij:2015yra} measurements \cite{He:2012zp,He:2018uey,Greljo:2018ogz,Babu:2018vrl,Gomez:2019xfw}. To address the viability of this explanation we examine the allowed parameter space in the $g_R-M_{W^\prime}$ plane with the specifics of Ref.~\cite{He:2018uey}.

The required $W^\prime$ would have a mass near 1~TeV, and Figure~\ref{f:wplimit} suggests that the largest coupling allowed in this region is approximately $g_R= 4$. For this parameter point the resonance width is $\Gamma_{W^\prime}/M_{W^\prime}=42\%$, still within the perturbative regime. The production of this resonance occurs from $cb$ annihilation and therefore it scales quadratically with the mixing angle $V^u_{Rtc}$. To obtain the constraints of the previous section we used  $V^u_{Rtc}=V_{cb}\approx 0.042$ from the ansatz of \cite{He:2009ie}. Clearly the constraints disappear if this mixing angle is smaller, but in that case the model is no longer a candidate to explain the charged B anomalies. The figure of merit to explain the latter is the product of the couplings $g_{bc}$  ($W^\prime b c$ coupling) and $g_{\tau\nu}$  ($W^\prime \tau \nu$ coupling) divided by the square of the $W^\prime$ mass. 

\begin{figure}[h!]
\centering{\includegraphics[width=.48\textwidth]{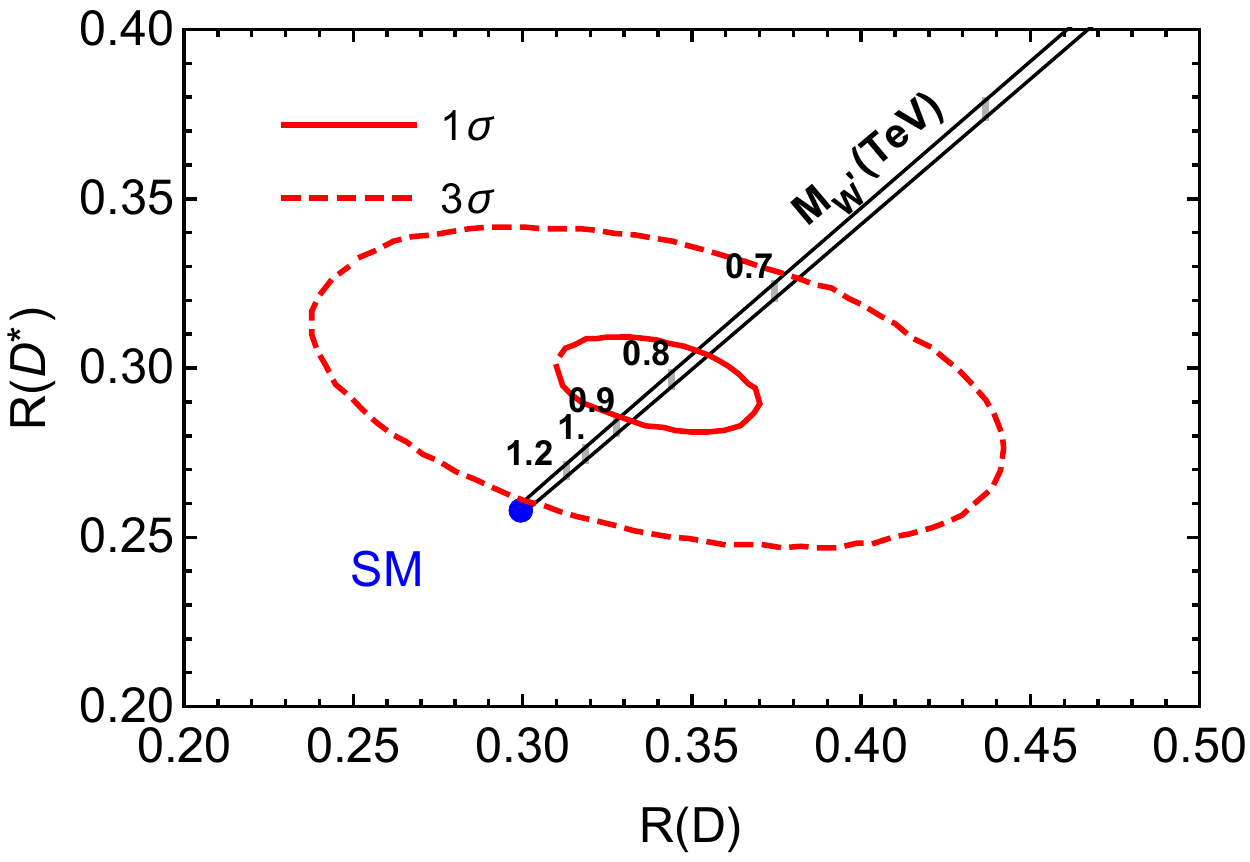}}
\caption{$R(D)$ vs $R(D^\star)$ HFLAV average from Spring 2019 \cite{Amhis:2016xyh} as red contours, the SM as a blue dot and the band predicted by the model of \cite{He:2018uey} for $g_R=4.0)$.}
\label{f:rdrds}
\end{figure}

Ref.~\cite{Greljo:2018tzh} argues that this explanation for $R(D)$ and $R(D^\star)$ is ruled out because fitting these quantities requires $|g_{bc}g^\star_{\tau\nu}|/M_{W^\prime}^2 =(0.6\pm 0.1)$~TeV$^{-2}$, which is ruled out by their recasting of the same CMS constraints we use in Figure~\ref{f:wplimit}. Our study of $pp \to W^\prime \to \tau N_R$ is in rough agreement with \cite{Greljo:2018tzh} but we find that this process is still consistent with explaining $R(D)$ and $R(D^\star)$ at the  $1\sigma$ level as can be seen in Figure~\ref{f:rdrds}. The figure shows the $1\sigma$ and $3\sigma$ contours from the HFLAV average from Spring 2019 \cite{Amhis:2016xyh} as well as the SM point. Superimposed on that result, the contribution of our $W^\prime$ is shown as the narrow black band. The width of the band is controlled by the smallness of the allowed $W-W^\prime$ mixing and the position along the band by the mass of the $W^\prime$ for $g_R=4$. The plot shows that the value $M_{W^\prime}=900$~GeV lies approximately on the $1\sigma$ HFLAV contour.
The point $g_R=4$, $M_{W^\prime}=900$~GeV, corresponds to $|g_{bc}g^\star_{\tau\nu}|/M_{W^\prime}^2 \approx 0.42$~TeV$^{-2}$ which is outside the  $1\sigma$ range quoted in Ref.~\cite{Greljo:2018tzh}. However, the HFLAV fits to $R(D)$ and $R(D^\star)$ (particularly for the former) are now lower than they were in 2018 and as a result, our model is compatible with these measurements, as illustrated in Figure~\ref{f:rdrds}.

\subsection{$K\to \pi \nu\bar\nu$}

These rare modes can receive contributions from a non-universal $Z^\prime$ both tree and one-loop level  as detailed in \cite{He:2004it,He:2018uey}. The tree-level contributions depend on FCNC  couplings of the $Z^\prime$ and are severely constrained by B mixing measurements. The one-loop contributions include penguins where the $Z^\prime$ couples to the top-quark and the sterile neutrino that are potentially large but quite model dependent. Their overall strength is determined by the ratio
\begin{eqnarray}
r_{Z^\prime}^2=\left(\frac{g_RM_Z}{g_LM_{Z^\prime}}\right)^2.
\end{eqnarray}
In order for these NP contributions to enhance the SM rates by factors of two, it was found in \cite{He:2018uey} that $r_{Z^\prime}$ needs to be of order 1. From Figure~\ref{f:zpcms} we see that for values of $g_R$ near its unitarity limit  $r_{Z^\prime}\lesssim 0.24$, 
about a factor of 4 smaller than the previous bound from LEP. With this new limit, it is no longer possible to have a large enhancement over the SM in the rare modes $K\to \pi \nu\bar\nu$ in this type of models.

\section{Summary and Discussion}

\begin{figure}[h!]
\centering{\includegraphics[width=.48\textwidth]{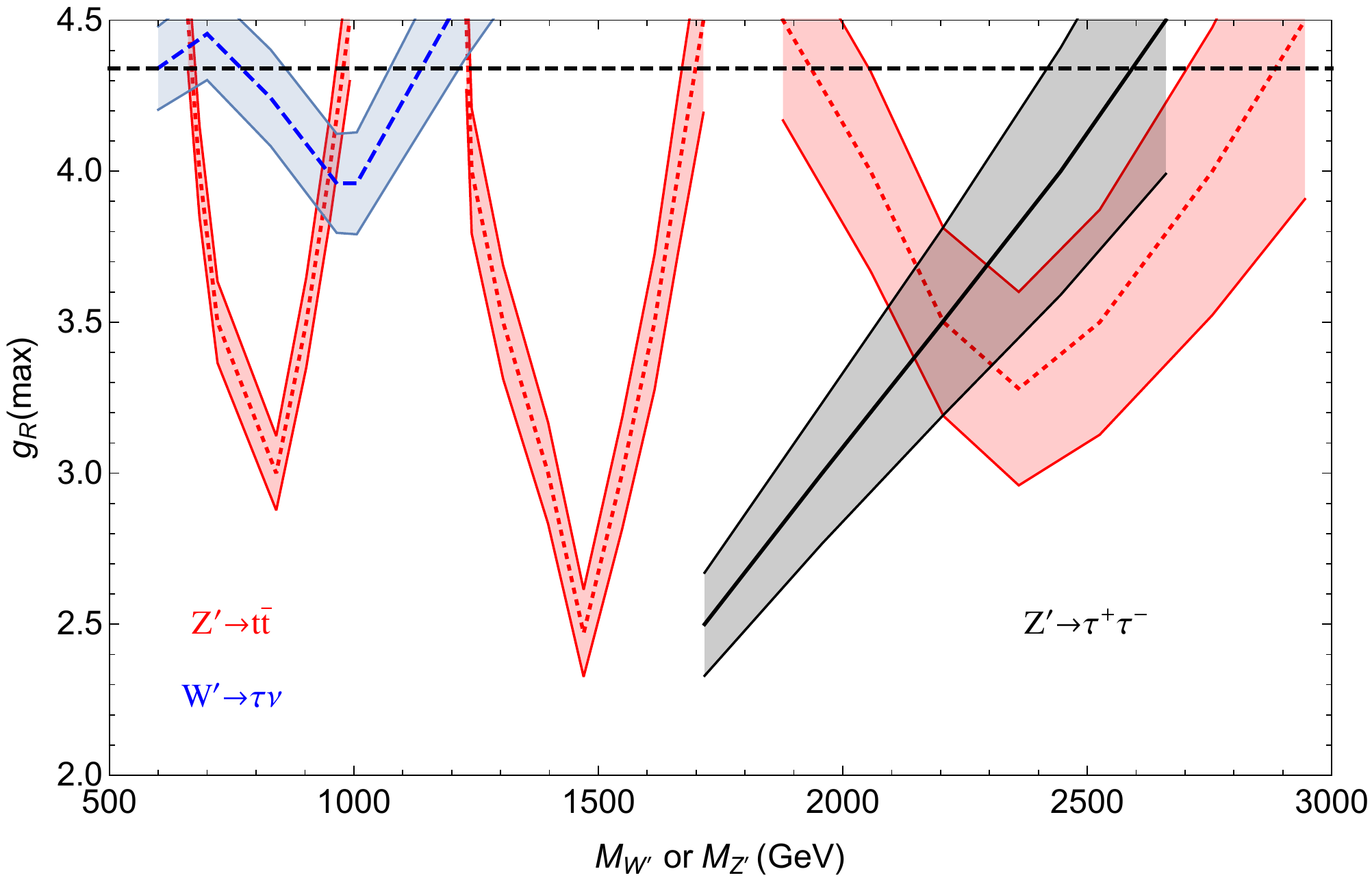}}
\caption{Largest allowed coupling $g_R$ as a function of $M_{W^\prime}$ from CMS studies of $pp\to \tau_h p_T^{\rm miss}$  (dashed blue line) compared to ATLAS studies of $pp\to t \bar{t}$ (red dotted line) and to CMS studies of $pp \to \tau\tau$ (solid black line). The dashed horizontal line marks the perturbative unitarity limit. }
\label{f:res}
\end{figure}

The constraints that can be placed on $W^\prime$ and $Z^\prime$ masses at the LHC by the different processes we considered in this paper are summarised in Figure~\ref{f:res}.
The  best constraint on the $W^\prime$ arises from the CMS studies of $pp\to \tau_h p_T^{\rm miss}$ and is shown as the dashed blue line in the figure.  Its coupling to third generation fermions is allowed to be as large as its perturbative unitarity limit for almost all values of $M_{W^\prime} \gtrsim 500$~GeV. In contrast the $Z^\prime$ is much more constrained. The ATLAS studies of $pp\to t \bar{t}$ are shown as the red dotted line in the figure and overall they indicate significant bounds on the strength of the coupling. The CMS studies of $pp \to \tau^+\tau^-$, produce a constraint  shown as a solid black line, and are even more restrictive, at least for $M_{Z^\prime} < 2$~TeV. Combined, these two studies push potential  $Z^\prime$ bosons with couplings to third generation fermions stronger than 5 times electroweak couplings beyond 2~TeV. In Figure~\ref{f:res} we have illustrated how the PDF uncertainties shown in Figure~\ref{pdfun} propagate to the constraints on coupling constants.

The $W^\prime$ is much less constrained than the $Z^\prime$ because its production requires at least one fermion from the first two generations (in our model a charm quark). This also makes the results dependent on right-handed quark mixing angles, and suggests that future LHC studies on double resonance production may be able to better constrain this type of  $W^\prime$. 

Considered independently, the LHC constraints on the $W^\prime$ still allow the light sterile neutrino explanation for $R(D^{(\star)})$, whereas those on the $Z^\prime$ rule out sizeable enhancements to $K\to \pi \nu \bar\nu$ rates in the same scenario. 

Within specific models the $W^\prime$ and  $Z^\prime$ masses are related, but this relation depends on parameters of the scalar sector. For the  models described in \cite{He:2002ha} the  large number of parameters  can be reduced for simplified cases as discussed in \cite{He:2003qv}. In that example, the vanishing of the gauge boson mixing parameters $\xi_W$ and $\xi_Z$ (at tree level) is accomplished with conditions on the vevs,  $v_2=0$ and $v_L = v_1\cot\theta_R$. Furthermore, the condition $v_R>>v_{L,1,2}$ separates the scales of symmetry breaking and makes the right handed gauge bosons much heavier than the $W$ and $Z$ but still approximately degenerate in mass. To split the $W_R$ and $Z_R$ masses one needs to introduce additional scalar representations such as triplets \cite{He:2002ha}. In the $g_R>>g_L$ limit, a $\Delta_R (1,1,3)(2)$ scalar with a vev will lead to  $M^2_{Z_R}/M^2_{W_R} = (v^2_R + 4 v^2_{\Delta_R})/(v^2_R + 2 v^2_{\Delta_R})$. However it will  also provide a Majorana mass to the neutrinos. As a  result, a large split between the $W_R$ and $Z_R$ masses achieved with $v_{\Delta_R}$ comparable to $v_R$ naturally leads to a heavy sterile neutrino which is then incompatible with a solution to $R(D^{(\star)})$. This  combination of conditions makes the light sterile neutrino explanation for $R(D^{(\star)})$ contrived but not impossible within these models, as the scalar sector can always be augmented.  For example a $\phi_{3/2} (1, 1, 3/2)(3)$ scalar, with a non-zero vev $v_{3/2}$ results in $M^2_{Z_R}/M^2_{W_R} =(v^2_R + 9 v^2_{3/2})/(v^2_R +  3 v^2_{3/2})$. If $v_{3/2}$ is of the same order as $v_R$, $M_{Z_R}$ can be much larger than $M_{W_R}$. In conclusion, unless one has a very detailed model in mind, the LHC constraints on $W^\prime$ and $Z^\prime$ masses should be viewed as independent.

\section*{Acknowledgements}

We are grateful to Ursula Laa and Bora Işıldak for useful discussions; Jorge Martin Camalich for clarifications on Reference \cite{Greljo:2018tzh}; and Christian Preuss for his help with Rivet. The work of GV and XGH was supported in part by the  Australian Research Council. XGH was supported in part by the MOST (Grant No. MOST 106-2112-M-002-003-MY3 ).

\appendix

\bibliography{wpzp}

\end{document}